\documentclass[twocolumn,showpacs,amsmath,amssymb]{revtex4}
\usepackage{graphicx}
\usepackage{bm}
\begin{document}
\title{The role of striction at magnetic and structural transitions in iron-pnictides}
\author{Victor Barzykin} 
\author{Lev P. Gor'kov}
\altaffiliation[Also at ]{L.D. Landau Institute for Theoretical Physics,
Chernogolovka, 142432, Russia}
\affiliation{National High Magnetic Field Laboratory,
Florida State University,
1800 E. Paul Dirac Dr., Tallahassee, Florida 32310 }

\begin{abstract}  
We discuss the role of striction in the intertwined magnetic and structural phase transitions in the 
underdoped iron-pnictides. The magneto-elastic coupling
to acoustic modes is then derived and estimated in framework of the multiband spectrum
for itinerant electrons with nesting features. We argue that the 1-st order
character of the magneto-elastic phase transition originates from the lattice instabilities 
near the onset of spin-density wave order introducing, thus, a shear acoustic mode as a new
order parameter. Taking non-harmonic termis in the lattice energy into account may 
explain the splitting of the structural and magnetic transitions in some oxypnictides. 
Fluctuations of the magnetic order parameter show up in the precursory temperature 
dependence of the elastic moduli.
\end{abstract}
\vspace{0.15cm}

\pacs{74.70.-b,74.20.-z,74.20.Rp,74.20.Mn}

\maketitle

Considerable efforts have been concentrated recently on studies of the interdependence for the 
antiferromagnetic (AFM) and the superconducting (SC) transitions  in the new iron-based 
compounds\cite{Kamihara}. Doping is known to destroy AFM and increase $T_C$ - the temperature 
for onset of SC\cite{huang}.

In this presentation we address the peculiarities and interrelations for the magnetic ($T_m$) 
and structural ($T_{str}$) transitions in the parent and underdoped iron-pnictides. 
For concreteness, we discuss below the properties of two systems only: the quarternary, 
REFeAsO (``1111''; RE stands for a rare earth), and the bi-layered, AFe$_2$As$_2$(``A22'', 
where A=Sr,Ca,Ba), materials. In both classes the intertwinned magnetic 
and the structural transitions of the weak 1-st order are observed in a temperature interval
of $\sim$ 100-200 K. In 1111's the structural change precedes the magnetic 
ordering\cite{huang,delacruz}, while in A22's magnetic order and change in the lattice 
symmetry occur simultaneously in the single 1-st order transition\cite{goldman}.

It is known that coupling to the lattice may transform a magnetically driven transition into 
the weak 1-st order transition accompanied by structural changes\cite{bean}.
In Ref. \cite{LPik} the problem has been rigorously
solved for the elastically isotropic solid, taking fluctuations into account. Unfortunately,
the method of Ref.\cite{LPik} does not apply to anisotropic materials, such as iron-pnictides 
with the tetragonal symmetry and layered structure. The solution of Ref.\cite{LPik}, however, 
undoubtedly contains the main physics for these phenomena. Correspondingly, below we simplify 
the approach of Ref. \cite{LPik} to model the magneto-elastic interactions (striction)
and phase transitions in the parent and underdoped FeAs-systems. 

The magnetic order in oxypnictides is built of the alternating (AFM) spins running along one 
of the lattice axis, while in the other direction the spins are ordered ferromagnetically\cite{delacruz}.
In the tetragonal
notations, the structure vectors are commensurate: $\bm{Q_1} = (0, \pi)$ and $\bm{Q_2} = (\pi,0)$.
With the local picture of interacting Fe spins and the two Heisenberg exchange interactions constants $J_1$ (nn) and $J_2$ (nnn),
such a ground state realizes itself at the inequality $J_2 > J_1/2$. Accompanying orthogonal lattice
distortions can be understood, as in Ref. \cite{bean}, in terms of a ``spin-Peierls'' effect, i.e.,
the variation of the exchange integrals at the lattice deformation\cite{yildirim}.

The new materials, however, are semimetals. Pnictides are better described
in terms of an itinerant scheme. The consensus is that the energy spectrum obtained in the 
``first principle'' calculations\cite{singh:du} presents a good starting point. According to 
\cite{singh:du}, the electron spectrum bears the multiband character: there are Fermi surfaces (FS) 
for the two hole-pockets (h-) at the 
$\Gamma$-point $(0,0)$ and two electronic pockets (e-) located in the tetragonal (unfolded)\cite{mazin}
reciprocal lattice at $(0,\pi)$ and $(\pi,0)$. An approximate nesting between the e- and h-pocket is
believed to be responsible for a spin-density wave (SDW) instability with
the two vectors, $\bm{Q_1}$ and $\bm{Q_2}$, mentioned above\cite{mazin}. The main features of this spectrum have
been reproduced in many numerical calculations for all classes of the new Fe-based
materials. In Ref. \cite{coldea} this spectrum has been directly observed for LaFePO in the 
de Haas-van Alphen (dHvA) experiments. Below we accept this model. With
$T_c \sim 50K$, $T_m \sim T_{str} \sim 100-200K$ and the Fermi energy $E_F \sim 0.1-0.2 eV$ for 
the pockets' sizes, the model is expected to allow a mean field treatment both for SC
and magnetic phenomena\cite{BGoxy}.

``First principle'' calculations provide a reasonable description of the ground state
properties, albeit some disagreements are not uncommon in the literature (see in Ref. \cite{mazin1}).
However, subtleties, especially in a vicinity of phase transitions, remain beyond
the reach of numerical analysis.

Among advantages of the "nesting" model\cite{mazin,KK,KM} is that its formalism is practically identical to the well-studied
BCS scheme for SC. It is a weak-coupling mean-field scheme if $T_{SDW} \ll E_F$. 
(Fluctuations become important only in a narrow temperature interval, $|\Delta T|/T_0 = Gi \ll 1$, as given 
by sort of a Levanyuk-Ginzburg criteria, $Gi$.) 
The Landau functional near the transition can be derived exactly as it has been
done for superconductivity near $T_c$\cite{gorkov}. 

Nevertheless we prefer not to write equations explicitly. Numerous unknown parameters 
that include parameters of the e- and h-pockets, interactions and six elastic moduli 
for the tetragonal symmetry of pnictides should make it useless. For the sake of transparency we 
hold, where possible, the discussion on the qualitative level.

\begin{figure}
\includegraphics[width=3.375in]{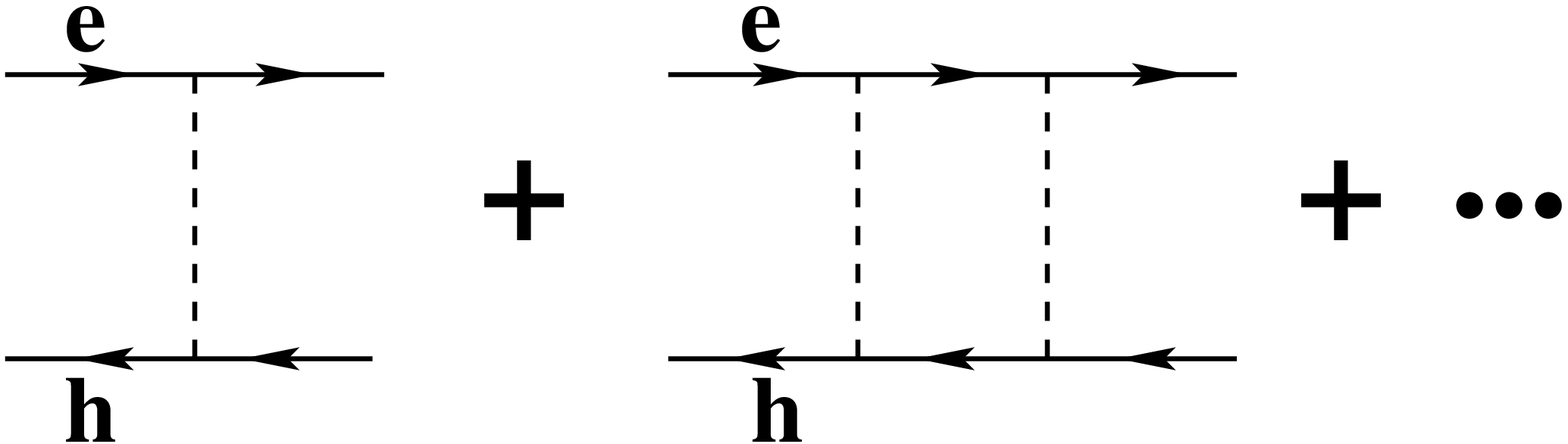}
\includegraphics[width=3.375in]{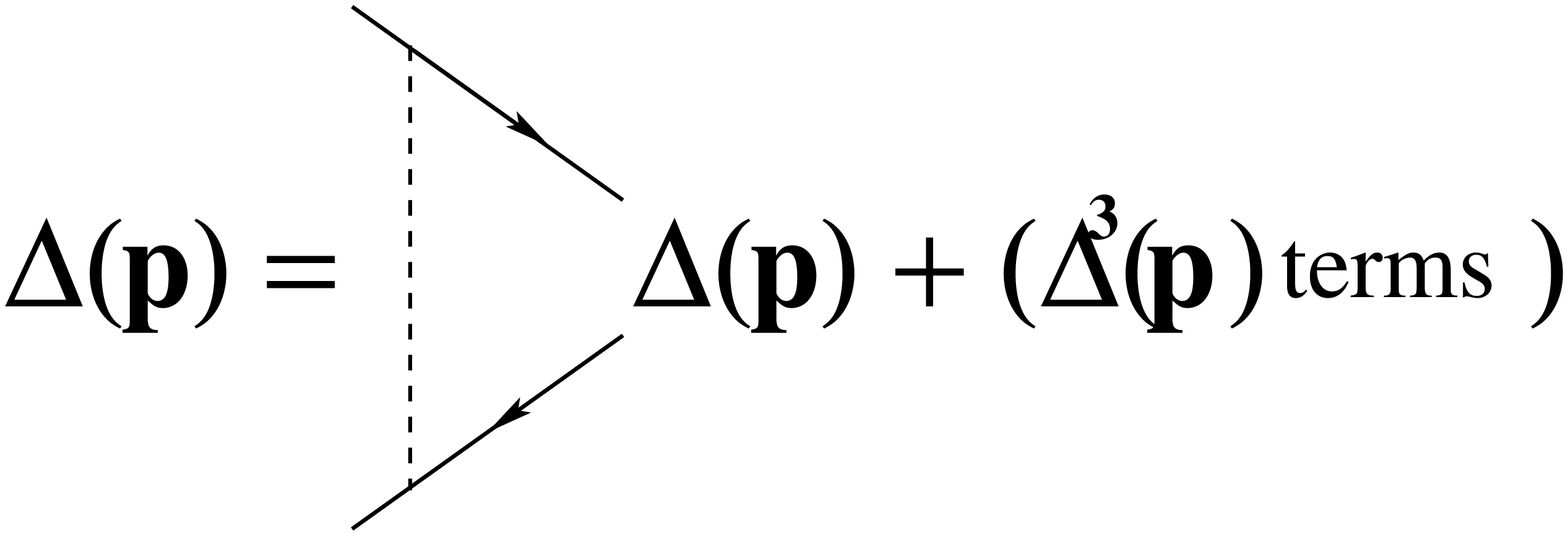}
\caption{(a) The logarithmic nesting model (b) Mean field equations near $T_c$.}
\label{fig1}
\end{figure}

The mathematical analogy of the model \cite{KK,KM} with the Cooper pairing in a BCS-like 
scheme stems from a logarithmic divergence for the scattering in the e-h-channel in Fig. \ref{fig1}a, which has the following form:
\begin{widetext}
\begin{equation}
V(\bm{p}-\bm{p'}) T \sum_{\omega_n} \int \frac{\nu(E_F) d \epsilon' d \Omega_{\bm{p'}}}{(i \omega_n - \epsilon_e')(i \omega_n - \epsilon_h')} =
V \nu(E_F) \ln{(\bar{\omega}/T, \delta)},
\label{nest}
\end{equation}
\end{widetext}
where
\begin{equation}
\epsilon_h = - \epsilon_e + \delta
\end{equation}
are e- and h- energies, while $\delta$ accounts for deviations from the ideal nesting.
With the notations, $\bm{\Delta_1}$ and $\bm{\Delta_2}$ for the triplet (SDW) order  parameters
corresponding to the structure vectors $\bm{Q_1} = (0, \pi)$ and $\bm{Q_2} = (\pi,0)$, the
equations for $\Delta$-s have the same structure as in SC theory\cite{gorkov}, as demonstrated in Fig. \ref{fig1}b.
Let $T_0$ be the ``bear'' transition temperature in the absence of coupling to the lattice. 
The Landau functional\cite{LL} near $T_0$ is of the form:
\begin{widetext}
\begin{equation}
\Omega = \int \frac{A \nu}{2}\, \left[ \tau \Delta^2 + b_0 \frac{\Delta^4}{T_0^2}\, + c_0 \xi_0^2  (\bm{\nabla}\bm{\Delta})^2 \right] d \bm{r},
\label{Lfunc}
\end{equation}
\end{widetext}
where $\nu \equiv \nu(E_F)$ is the characteristic density of states (DOS). $A$, $c_0$, $b_0$ are constants, and
$\tau = (T-T_0)/T_0$; $\xi_0 = v_F/2 \pi T_0$. The functional Eq.(\ref{Lfunc}) is the starting
point in the general theory of the second order phase transitions\cite{LL,pokrovskii:book}. To account
for interactions with the lattice, one adds to Eq.(\ref{Lfunc}) a coupling to the lattice of the form:
\begin{equation}
H_{str} = -q \int \hat{u}(\bm{r}) \hat{\Delta}(\bm{r})^2 d \bm{r},
\label{inter}
\end{equation}
and the elastic energy itself that we also write schematically as:
\begin{equation}
H_{el} = \int K \hat{u}^2 d \bm{r},
\label{struc}
\end{equation}
where $K$ is an elastic modulus (such as the bulk modulus, for instance);
the notation $\hat{u}$ stands for components of the strain tensor, 
\begin{equation}
u_{ik} = \frac{1}{2} \left( \frac{\partial u_i}{\partial x_k}\, + \frac{\partial u_k}{\partial x_i}\, \right).
\end{equation}
For magnetic phenomena, Eq.(\ref{inter}) describes magneto-elastic interactions and $q$ is called the striction constant.
The total Gibbs energy near the phase transition is
\begin{equation}
\Phi = \Omega(\tau) + H_{el} + H_{str},
\label{pot}
\end{equation}
In the itinerant model of Eq.(\ref{nest}) the parameters $\Delta_1$ and $\Delta_2$ are built
on the Bloch wavefunctions. These spatial features present in the SDW parameters immediately lead to the 
quadratic coupling between $\Delta$-s and acoustic degrees of freedom ($\bm{Q} = 0$) given by  Eq.(\ref{inter}). An estimate
of the striction constants in Eq.(\ref{inter}) follows directly from the logarithmic contribution of Eq.(\ref{nest}).
Indeed, elastic deformations change the parameter $\delta$
in Eq.(\ref{Lfunc}) that controls the degree of the nesting:
\begin{equation}
\delta \equiv \delta(u) = \delta_0 + \lambda \hat{u}
\label{delt}
\end{equation}

From Eq.(\ref{nest}) and Fig. \ref{fig1}b it is seen that
\begin{equation}
\tau \rightarrow \tau + \lambda \frac{\hat{u}}{T_0}\,
\label{subst}
\end{equation}
Making use of Eq.(\ref{Lfunc}), we estimate
\begin{equation}
q \sim \frac{\nu \lambda}{T_0}\, \sim \nu \left( \frac{E_F}{T_0}\, \right)
\label{qesti}
\end{equation}
(taking $\lambda$ of order of characteristic atomic energy, $E_F$).

Return to Eqs. (\ref{inter})-(\ref{pot}) and assume homogeneous $\hat{u}$. 
Minimizing $\Phi$ over $\hat{u}$, we obtain:
\begin{equation}
K \hat{u} = q \left(\frac{\int \Delta^2 d \bm{r}}{V}\, \right)
\label{keffe}
\end{equation}
(with $V$ for the volume). The effective Landau functional $\tilde{\Omega}$ is
\begin{equation}
\tilde{\Omega}_{min} = \Omega(\tau) - \left(\frac{1}{2 K}\, \right) \left[q \frac{\int \Delta^2(\bm{r}) d \bm{r}}{V}\, \right]^2
\label{om}
\end{equation}
In the mean field $\Delta^2(\bm{r}) = const$, and Eq.(\ref{om}) reproduces the result of Ref.\cite{bean}: the
functional, $\Omega$, acquires a negative contribution to the biquadratic term in Eq.(\ref{Lfunc}).
Should the total be negative, the transition becomes 1-st order. (Higher order terms in 
the expansion Eq.(\ref{Lfunc}) become necessary.)

In the presence of an external homogeneous deformation, $\hat{u}_{ext}$
substitution into Eq.(\ref{subst}), $\tau \longrightarrow \tau + \lambda (\hat{u}_{ext}/T_0)$, leads to: 
\begin{equation}
\Omega_{eff} = \Omega\left(\tau + \lambda \frac{\hat{u}_{ext}}{T_0}\, \right).
\end{equation}
Differentiating, we find:
\begin{equation}
\frac{\partial^2 \Omega_{eff}}{\partial \hat{u}_{ext}^2}\, = 
\frac{\lambda^2}{T^2}\, \frac{\partial^2 \Omega}{\partial \tau^2}\, = - \frac{\lambda^2 C(\tau)}{T_0}\,,
\label{secder}
\end{equation}
where $C(\tau)$ is the specific heat.  According to Eq.(\ref{secder}), already in the Landau mean field
approach the jump $\Delta C$  in the specific heat at the transition is accompanied by a negative step-like
change of elastic moduli. The variation is of the atomic order. Therefore the lattice would become 
unstable, and the 2-nd order character of a transition changes to the 1-st order one.

In Refs.\cite{LPik,Rice}, the attention has been drawn to the fact that the specific heat $C(T)$ in Eq. (\ref{secder})
actually has a singularity at $T_0$:
\begin{equation}
C(T) \propto |\tau|^{- \alpha}.
\label{sph}
\end{equation}
($\alpha$ is the so-called scaling index for the specific heat, $C(T)$.)
 As to
the order parameter, it appears below $T_0$ (see Ref.\cite{LL}, \S 148, p. 484): 
\begin{equation}
\Delta(T) \propto (- \tau)^{\beta} (\tau < 0)
\label{opar}
\end{equation} 

Therefore an elastic instability is ubiquitous for 
any magnetic transition. In the model chosen above the instability 
occurs inside a narrow temperature interval controlled by the Levanyuk-Ginzburg
parameter. Provided that layered pnictides can be treated, in the first approximation, as 
two-dimensional (2D), we write:
\begin{equation}
\left| \frac{\Delta T}{T_0}\, \right| \equiv G^{2D}_i \sim \left(\frac{T_{SDW}}{E_F}\, \right) \ll 1.
\label{2dcr}
\end{equation}
By the order of magnitude, the same parameter also controls the width of the hysteresis.

Note that even for homogeneous
lattice deformations the magneto-elastic coupling, Eq.(\ref{inter}), leads to Eq.(\ref{om}), 
with the new biquadratic term that describes 
non-local interactions for the order parameter $\Delta$. In Ref. \cite{LPik}
inhomogeneous fluctuations result in the cancellation of non-local terms Eq.(\ref{om}) for
the isotropic (liquid) media. However, the non-local coupling terms 
are always present for any anisotropic solid. As it was mentioned above, the problem
could be rigorously solved in Ref\cite{LPik} only for the elastically isotropic solid (i.e., for a solid 
characterized by the bulk and shear moduli). In all other cases the 
non-local terms Eq.(\ref{om}) that come about due to exchange by acoustic phonons become strongly anisotropic.
We now simplify the model \cite{LPik} by restricting our consideration to homogeneous lattice fluctuations.
Inhomogeneous fluctuations should not qualitatively change the physics of the problem.

With this in mind, one can rewrite Eq.(\ref{Lfunc}) in the form:
\begin{equation}
\Phi = \frac{K}{2}\, \hat{u}^2 + \Omega\left(\tau + \frac{\lambda}{T_0}\, \hat{u} \right) - \hat{\sigma} \hat{u},
\label{phi}
\end{equation}
where  $\Omega\left(\tau + \frac{\lambda}{T_0}\, \hat{u} \right)$ is the exact functional
Eq.(\ref{Lfunc}), i.e., the functional that would describe the second order phase transition  
driven by the order parameter $\Delta$ neglecting striction effects. In particular, it has the
contribution, $\Omega_{sing}$, responsible for the singular behavior of the specific heat,
$C(T)$, Eq.(\ref{sph}). The term $- \hat{\sigma} \hat{u}$ stands for applied 
external stress ($\hat{\sigma}$ is a proper component of the stress tensor; this term
is needed at calculations of the temperature behavior of elastic moduli
across the transition.) 

In the presence of non-zero stress, $\hat{\sigma} \neq 0$, fluctuations take place around the new
equilibrium point, $\hat{u}_{ext}$:
\begin{equation}
\hat{u}_{ext} = \frac{\hat{\sigma}}{K}\,
\label{uext}
\end{equation}
Rewriting in Eq. (\ref{phi}) $\hat{u} \rightarrow \hat{u} + \hat{u}_{ext}$, we obtain:
\begin{equation}
\Phi = - \frac{\hat{\sigma}^2}{2 K}\, + \frac{K}{2}\, \hat{u}^2 + 
\Omega\left(\tau + \frac{\lambda}{T_0}\, \hat{u} + \frac{\lambda \hat{\sigma}}{T_0 K}\, \right).
\label{phi1}
\end{equation}
The contribution $\lambda \hat{\sigma}/(T_0 K)$ in the argument of $\Omega(x)$, Eq.(\ref{phi1}),
shifts the transition temperature $T_0$ under applied stress,
\begin{equation}
T_0(\hat{\sigma})= T_0 - \frac{\lambda \sigma}{K}\,.
\end{equation}
The Gibbs energy Eq.(\ref{phi1}) must be minimized over fluctuations of (homogeneous) deformations, $\hat{u}$:
 \begin{equation}
\frac{\delta \Phi}{\delta \hat{u}}\, = K \hat{u} +\frac{\lambda}{T_0}\, \Omega'(x) = 0,
\label{phi2}
\end{equation}
where
\begin{equation}
x = \tau + \frac{\lambda}{T_0}\, u + \frac{\lambda \sigma}{T_0 K}\,.
\label{xdef}
\end{equation}

One
may easily recognize in Eqs (\ref{phi1}) and (\ref{phi2}) the analogy to equations
in Ref. \cite{LPik} describing the transitions in the elastically isotropic solid body.

To be now more specific, note that the elastic energy of homogeneous deformations for 
the tetragonal lattice can be rewritten in terms of the even irreducible representation of the $D_{4h}$ group:
\begin{equation}
A_{1g}: u_{xx} + u_{yy}, u_{zz}; B_{1g}: u_{xx} - u_{yy}; B_{2g}: u_{xy}; E_g: u_{zx}, u_{zy}.
\label{groups}
\end{equation}

In principle, any distortion Eq.(\ref{groups}) perturbs the electronic spectrum
and, hence, affects the nesting features. Therefore, the number of the 
independent striction constants, $q$-s, in Eq.(\ref{inter}) (or $\lambda$-s in Eqs (\ref{delt}), (\ref{subst}))
may be large. In view that the layered character of pnictides makes them close to two-dimensional systems, the
nesting parameter $\delta$ in Eq.(\ref{delt}) is controlled mainly by the strain components in Eq.(\ref{groups}) that
do not have the z-indicies. Rewriting the 2D-part of the elastic energy Eq.(\ref{struc}) in the form
\begin{equation}
H_{el} = \frac{K}{2}\, (u_{xx} + u_{yy})^2 + \frac{\mu_1}{2}\, (u_{xx} - u_{yy})^2 + 2 \mu_2 u_{xy}^2
\label{el1}
\end{equation}
with three independent moduli according to Eq.(\ref{groups}) and three magneto-elastic constants $q_i$ (or $\lambda_i$ in Eqs.(\ref{delt}), (\ref{subst}))
 and taking into account the uniaxial symmetry
along $(0, \pi)$ or $(\pi, 0)$ directions for the parameters $\Delta_1$ and $\Delta_2$, we simplify the problem further and rewrite
Eq.(\ref{subst}) as
\begin{equation}
\tau \rightarrow \tau + \frac{\lambda_+}{T_0}\, (u_{xx} + u_{yy}) + \frac{\lambda_-}{T_0}\, (u_{xx} - u_{yy})
\end{equation}
The second term is responsible for the orthorombic deformation of the lattice.

For the actual calculations one would need the expression for $\Omega(\tau)$. In our case the fluctuations are strong only in
a narrow vicinity of the phase transition, and we expect that the 1-st order transition occurs inside the same interval. 
Therefore
only the \textit{singular} part, $\delta \Omega_{sing} (\tau)$, is of the importance. We limit ourselves by the 
first fluctuation correction to the mean field $\Omega$, which we calculate in exactly the same manner as in Ref.\cite{LL} (see \S 147, problem, p. 482). 
The minor difference is that for the strongly anisotropic pnictides one needs to introduce in Eq.(\ref{Lfunc}) the in-plane and out-of-plane coherence 
lengths, $\xi_{0 \parallel} \gg \xi_{0 \perp}$. After an elementary calculation it follows:
\begin{equation}
\delta \Omega_{sing}(\tau) = - const (T_0 \xi_{0 \parallel}^{-2} \xi_{0 \perp}^{-1} ) |\tau|^{3/2} \equiv - B |\tau|^{3/2},
\label{singular}
\end{equation}
where $const$ is a numeric factor that depends on the model details. (Strictly speaking, the values of $B$ in Eq.(\ref{singular}) 
differ by factor $2^{3/2}$ on the two sides of $T_0$\cite{LL} (see \S 146, footnote on p. 475).) For the singularity in the specific heat, Eq.(\ref{singular}) gives:
\begin{equation}
\delta C (\tau) = \frac{3}{4 T_0}\, B |\tau|^{-1/2}
\label{Csing}
\end{equation}
Comparing Eq. (\ref{Csing}) with the normal specific heat, $C_n \sim \nu T_0$ , leads to the criterion:
\begin{equation}
\frac{B}{\nu T_0^2} |\tau|^{-1/2} \equiv \alpha |\tau|^{-1/2} \ll 1
\label{cri}
\end{equation}
(In the strictly 2D limit ($\xi_{\perp0} \rightarrow \infty$) one would obtain the criteria Eq.(\ref{2dcr})).
With Eqs(\ref{secder}) and (\ref{xdef}) one finds that corrections to the elastic moduli of Eq.(\ref{el1}) 
become strong in the same order of magnitude temperature 
interval Eq.(\ref{2dcr}) (or Eq.(\ref{cri})) if both $K$ and $\mu$ are of the atomic scale.

To demonstrate the emergence of the 1-st order transition from Eqs (\ref{phi1}),(\ref{phi2}), assume, for simplicity sake, 
$\alpha = 1/2$ in Eq.(\ref{sph}),  i.e., extrapolate Eq.(\ref{singular})
over the whole fluctuation interval. From Eq.(\ref{phi2}) (for each $u_{+,-}$) it follows:
\begin{equation}
K u_+ = \frac{3}{2}\, B \frac{\lambda_+}{T_0}\, \frac{x}{|x|}\, |x|^{1/2}; \ \ u_- \mu_1 = \frac{3}{2}\, B \frac{\lambda_-}{T_0}\, \frac{x}{|x|}\, |x|^{1/2}.
\label{Keq}
\end{equation}
With the notations
\begin{equation}
\alpha_+ = \frac{3}{2}\, B \frac{\lambda_+^2}{T_0^2}\,; \ \ \alpha_- =  \frac{3}{2}\, B \frac{\lambda_-^2}{T_0^2}\,
\end{equation}
\begin{equation}
\frac{\lambda_+ u_+}{T_0}\, = \rho, \ \ \frac{\lambda_+ u_+}{T_0}\, = \mu, \ \ x \equiv \tau + \rho + \mu,
\label{not2}
\end{equation}
we rewrite Eqs.(\ref{Keq}) as
\begin{equation}
\frac{1}{\alpha_+} \rho = \frac{x}{|x|}\, |x|^{1/2}; \frac{1}{\alpha_-} \mu =  \frac{x}{|x|}\, |x|^{1/2}
\end{equation}
or
\begin{equation}
\alpha_+^{-1} \rho = \alpha_-^{-1} \mu; \ \ x = \tau + \rho \left(1 + \frac{\alpha_-}{\alpha_+}\, \right) = \tau + v.
\end{equation} 
Finally, with 
\begin{equation}
v = (\alpha_+ + \alpha_-) \frac{x}{|x|}\, x^{1/2}
\end{equation}
we arrive to the single equation
\begin{equation}
x = \tau + (\alpha_+ + \alpha_-) \frac{x}{|x|}\, x^{1/2}
\label{solut}
\end{equation}
Eq.(\ref{solut}) like Eq.(13) of Ref.\cite{LPik}, reveals the typical
features of a 1-st order transition: at small enough $\tau$ there are three solutions for $x$. 
If coefficients $B$ in Eq.(\ref{singular}) are equal on both sides of $T_0$,
the transition takes place at $\tau = 0$, where 
\begin{equation}
x_+ = - x_- = (\alpha_+ + \alpha_-)^2
\end{equation}
determines the jumps of $u_+,u_-$ at the transition ( note that at negative $x_- \neq 0$ the driving 
parameter, according to Eq.(\ref{opar}), is finite). The area of hysteresis is determined by the equation:
\begin{equation}
2 = (\alpha_+ + \alpha_-) |x|^{-1/2}
\label{hyst}
\end{equation}

These results qualitatively agree with the observed simultaneous onset both of orthorombic distortions and 
the  ``stripe'' SDW order in the iron system A22\cite{goldman}. The distortions Eq.(\ref{not2}) just accompany the magnetic
transition. In case of the ``1111'' class the temperature, $T_{str}$, for the onset of the orthorombic deformation precedes
onset of the magnetic order at $T_m$\cite{delacruz}. The two temperatures are rather close: $\Delta T = T_{str} - T_m > 0$
is of order of $10-20 K$, so that by order of magnitude $\Delta T/T_{str} \sim 0.1$ falls into the range of Eq. (\ref{2dcr}) 
with $E_F \sim 0.1 - 0.2 eV$ taken for the pockets' depths obtained in ``first principle'' calculations\cite{singh:du,sadovskii}.
Such closeness seen among most of REFeAsO (e.g. compare La\cite{delacruz} and Nd \cite{chen}) is strongly in favor that the 
structural transition is directly related to the magnetic instability. In the language of local Fe-spins the attempt was
made in \cite{kivelson} to ascribe the temperature interval separating $T_{str}$ and $T_m$ to appearance of a ``nematic phase''
that comes about due to strong spin fluctuations above $T_m$. We suggest that both transitions have the common origin and come
about as the result of the lattice instabilities caused by striction. Indeed, according to Eq.(\ref{secder}), striction triggers softening of elastic moduli
as temperature approaches the transition interval Eq.(\ref{2dcr}) or Eq.(\ref{cri}). Assume that it takes place more strongly for modulus $\mu_1$, in Eq.(\ref{el1}). 
Recall that the orthorombic distortion is the symmetry change by itself and for the tetragonal lattice is characterized by the symmetry 
parameter $u_- = u_{xx} - u_{yy}$. So far as the dependence $\delta \Omega_{sing} (\tau + (\lambda_-/T_0) u_-)$ on $u_-$ is the only form of
the nonlinear elastic energy, the above analysis applies. However, when the renormalized modulus, $\mu_{1 eff}$, becomes small,
other non-linear terms ever present in the lattice must be also taken into account. As the result, the Landau functional 
for the parameter $u_- = u_{xx} - u_{yy}$ also depends on 
those contributions.  The higher order terms in $u_-$ becoming important
when renormalized $\mu_{1 eff}$ is small nearby magnetic $T_0$. The mean field treatment of the new symmetry parameter, $u_-$, could then be
applied in the usual way for the 2-nd order tetra-ortho transition at $T_{str}$. 

Experimentally, the structural distortions at $T_{str}$ appear in the weak 1-st order transition\cite{delacruz}. That last result immediately follows 
from the fact that by symmetry the orthorombic transition in the tetragonal lattice infers its \textit{own quadratic striction}.
Indeed, the following cubic terms are allowed in the tetragonal lattice by the symmetry reasons:
\begin{equation}
q_{\parallel} u_+ u_-^2 + q_{\perp} u_{zz} u_-^2
\label{cub1}
\end{equation}
in addition to the elastic terms:
\begin{equation}
\frac{K}{2}\, u_+^2 + \frac{K_{\perp}}{2}\, u_{zz}^2
\label{cub2}
\end{equation}
In terms of our model Eqs (\ref{nest}),(\ref{Lfunc}),(\ref{pot}) the fact that $T_m < T_{str}$ should mean that the non-zero
distortion $u_-$ below $T_{str}$ makes the nesting conditions Eq.(\ref{nest}) worse. 

Let's add a few final comments to the analyses above. The order of magnitude estimate of the
in-plane lattice distortions at low enough temperatures follows from Eqs (\ref{qesti}),(\ref{keffe}):

\begin{equation}
u \sim \frac{q}{K}\, \sim \frac{\nu E_F}{K T_0}\, T_0^2 \sim \frac{T_0}{E_F}\, \sim 10^{-2} - 10^{-3}.
\label{defes}
\end{equation} 
that agrees well with the experimental data of Ref.\cite{fratini,rotter,mcguire}. The hysteresis, $\Delta T$, from 
Eq.(\ref{hyst}) agrees by the order of magnitude with data \cite{goldman}

There are no experimental data for the volume change, $\Delta V/V$. The lattice deformation along the c-axis, in
principle, could be obtained from Eqs(\ref{cub1}) and (\ref{cub2}) together:
\begin{equation}
u_{zz} \sim \frac{q_{\perp}}{K_{\perp}}\, u_-^2.
\label{uzz}
\end{equation}
In Eq.(\ref{uzz}) both $q_{\perp}$ and $K_{\perp}$ are expected to be small in a layered material. These parameters remain unknown in the oxypnictides. 

Eq.(\ref{secder}) together with Eq.(\ref{singular}) provides another observable feature: at temperatures above $T_0$ the fluctuation corrections to the elastic moduli
should behave as
\begin{equation}
\sim (T - T_0)^{- 1/2},
\label{expo1}
\end{equation}
while in the 2D limit of isolated planes
\begin{equation}
\sim (T - T_0)^{-1}.
\label{expo2}
\end{equation}
The data\cite{mcguire} do not allow to distinguish between the exponents in Eqs(\ref{expo1}), (\ref{expo2}). 

To summarize, in the frameworks of multiband electronic spectrum with nesting features the theoretical scheme is
elaborated to treat striction in iron-pnictides. Magneto-elastic coupling changes the second order character of magnetic transition.
The transition becomes of the 1-st order. Whether the transition bears strong or weak 1-st order character
may depend on details. Provided that $T_0 \ll E_F$, the weak 1-st order transition is predicted. The
model, when applied to the layered FeAs systems, leads to estimates of the correct order of magnitude. 
Discontinuities of all parameters at the transition are due to lattice instabilities.  
Magneto-elastic interactions may split
the magnetic (SDW) transition at $T_m$ and the orthorombic deformations at $T_{str} > T_m$. The model predicts a noticeable
precursory temperature dependence above the transition temperature in the elastic moduli.

The obtained results are in good qualitative agreement with peculiarities of the phase diagram of new parent or
underdoped FeAs materials well above the temperature of superconducting transition.

The authors are thankful to Z. Fisk, A.Gurevich, J. Lynn, A. Migliori, V. Pokrovsky, I. Liuksyutov, D. Khmelnitsky  and  M. Sadovskii 
for helpful discussions. This work was supported by NHFML through the NSF
Cooperative agreement No. DMR-008473 and the State
of Florida.

\end{document}